\documentclass[aps,pra,twocolumn,superscriptaddress]{revtex4-2}
\usepackage{amsmath}
\usepackage{amsfonts}
\usepackage{bm}
\usepackage{braket}
\usepackage{graphicx}
\usepackage{verbatim}
\usepackage{siunitx}
\usepackage{units}
\usepackage{color}
\usepackage{xcolor}
\usepackage{dsfont}

\usepackage{tikz}
\newcommand*\circled[1]{\tikz[baseline=(char.base)]{
    \node[shape=circle,draw,inner sep=0.5pt] (char) {#1};}}
\newcommand*\dashcircled[1]{\tikz[baseline=(char.base)]{
        \node[shape=circle,draw,inner sep=0.5pt,style=densely dashed] (char) {#1};}}
\usepackage{hyperref}
\graphicspath{{figures/}}
\begin{document}

\title{Sparse interferometry for measuring multiphoton collective phase}
\author{Jizhou Wu}
\email{wujizhou@mail.ustc.edu.cn}
\affiliation{Shanghai Branch, National Laboratory for Physical Sciences at Microscale,
University of Science and Technology of China, Shanghai 201315, China}
\affiliation{%
CAS Center for Excellence and Synergetic Innovation Center in Quantum Information and Quantum Physics, University of Science and Technology of China,
Hefei, Anhui 230026, China%
}

\author{Barry C. Sanders}
\email{bsanders@ustc.edu.cn}
\affiliation{%
Shanghai Branch, National Laboratory for Physical Sciences at Microscale,
University of Science and Technology of China, Shanghai 201315, China%
}
\affiliation{%
CAS Center for Excellence and Synergetic Innovation Center in Quantum Information and Quantum Physics, University of Science and Technology of China, Hefei, Anhui 230026, China%
}
\affiliation{%
Institute for Quantum Science and Technology, University of Calgary, Alberta, Canada T2N 1N4%
}

\begin{abstract}
A multiphoton collective phase is a multiphoton-scattering feature
that cannot be reduced to a sequence of two-photon scattering events,
and the three-photon ``triad phase'' is the smallest nontrivial example.
Observing a higher-order collective phase is experimentally challenging,
and only triad and four-photon tetrad collective phases have been observed.
We introduce a scheme to make higher-order multiphoton collective-phase observations feasible by designing a
sparse interferometer,
which significantly reduces complexity compared with the current best scheme for observing a multiphoton collective phase.
Specifically, our scheme reduces the optical depth from logarithmic to constant and reduces the number of beam splitters from~$O(n\log n)$ to linear scaling
with respect to the collective-phase order~$n$.
As constant depth reduces loss and dispersion to a fixed rate regardless of collective-phase order,
a major obstacle to observing large-scale collective phases is removed.
\end{abstract}
\maketitle

\section{Introduction}
In the Hong-Ou-Mandel~(HOM) effect,
a celebrated example of two-photon interference,
two identical single photons arrive at distinct ports of a balanced (50:50) beamsplitter~(BS) with a controllable relative-time delay,
and quantum interference is manifested as an output coincidence-rate dip with the dip minimum occurring
when the time delay decreases to zero~\cite{Hong1987Phys.Rev.Lett.}.
The relationship between the photons' distinguishability, 
which is adjusted by the time-delay mechanism in the Hong-Ou-Mandel effect, 
and the controllable two-photon coincidence
makes two-photon interference an excellent tool to characterize single-photon sources,
such as spontaneous parametric downconversion~\cite{Riedmatten2003Phys.Rev.A,Kaltenbaek2006Phys.Rev.Lett.,Mosley2008Phys.Rev.Lett.,Halder2008NewJ.Phys.,Aboussouan2010Phys.Rev.A,Tanida2012Opt.Express} and quantum dots~\cite{Santori2002Nature,Sanaka2009Phys.Rev.Lett.,Ates2009Phys.Rev.Lett.,Flagg2010Phys.Rev.Lett.,He2013Nat.Nanotechnol.,Wei2014NanoLett.,Unsleber2015Phys.Rev.B,Wang2016Phys.Rev.Lett.,Ding2016Phys.Rev.Lett.,Senellart2017NatureNanotech.},
and reconstruct the scattering matrix of linear optical interferometers~\cite{Laing2012ArXiv12082868Quant-Ph,Rahimi-Keshari2013Opt.Express,Dhand2016J.Opt.,Tillmann2016J.Opt.}.

When generalized to multiphoton (more-than-two-photon) quantum interference with controllable mutual distinguishability,
a multiphoton collective phase (MCP),
which is a geometric phase in photonic internal space,
arises.
An MCP cannot be reconstructed from two-photon interference~\cite{Shchesnovich2018Phys.Rev.A}
and reveals facets of multiphoton distinguishability.
To minimize the phase effect coming from the interferometer and focus only on the collective phase from multiple photons,
experiments that manifest the collective phase are usually done with balanced multiport interferometers
(with ``balanced'' referring to a single photon entering any input port having an equal probability of being detected at any output port).

Three- and four-photon interferometers
(known as tritters~\cite{Zeilinger1993QuantumcontrolandmeasurementElsevier,Campos2000Phys.Rev.A}
and quitters and quarters~\cite{Greenberger1993Phys.Today,Mattle1995Appl.Phys.B},
which are balanced six-port and eight-port interferometers, respectively)
have been demonstrated to be able to observe MCPs~\cite{Menssen2017Phys.Rev.Lett.,Jones2020Phys.Rev.Lett.}.
Practically, to measure an $n$-photon MCP with a balanced multiport interferometer comprising $O(n\log n)$ optical elements with fixed loss~$\gamma$ per element,
the optimal optical depth
(the largest number of optical elements along the transmission path)
is~$O(\log n)$~\cite{Barak2007J.Opt.Soc.Am.B}.
The resultant photon loss rate is~$\gamma^{O(\log n)}$.

For multiphoton interference in the balanced multiport,
MCPs of all orders manifest in the interference.
Thus one needs to make pairs of photons totally distinguishable with the help of photonic internal degrees of freedom
such as polarization
(corresponding to a two-dimensional Hilbert space)
or time delay
(corresponding to an infinite-dimensional Hilbert space),
to eliminate lower-order MCPs,
thereby making the highest-order MCP prominent~\cite{Shchesnovich2018Phys.Rev.A}.
Specifically, we solve the problem of whether all lower-order MCPs can be eliminated with optical depth less than~$O(\log n)$.

We solve this problem,
in the positive,
by sparsifying the interferometer so that all MCPs are zero for fewer than~$n$ photons.
During the procedure of sparsification,
we need to increase the mode number of the interferometer to ensure that
the interferometer is lossless and described by a unitary matrix.
Then to use fewer optical elements,
we solve for the minimum mode number required in our procedure of sparsification.
With the help of graph theory,
we propose a~$2n$-mode sparse interferometer,
which turns out to deliver the desired MCP using only
constant,
rather than logarithmic,
optical depth.
The number of beam splitters is reduced from $O(n\log n)$ to $O(n)$.

Our results are important 
because we provide an approach to multiphoton state characterization that requires fewer resources 
than either full tomography 
or the fully connected interferometer method, 
while providing all necessary information to predict multiphoton-interference behavior. 
Thus, our results can be useful for a range of multiphoton interference experiments 
such as for boson sampling~\cite{Aaronson2013TheoryComput.,Wang2019Phys.Rev.Lett.b,Wang2018Phys.Rev.Lett.a,Wang2017Nat.Photonics,He2017Phys.Rev.Lett.,Carolan2014Nat.Photonics,Tillmann2013Nat.Photonics,Spring2013Science,Crespi2013Nat.Photonics,Broome2013Science}
and quantum metrology~\cite{Motes2015Phys.Rev.Lett.,Su2017Phys.Rev.Lett.}.

This paper is organized as follows.
In Sec.~\ref{sec:background} we provide a detailed background of MCPs
plus basic notation and a theoretical framework of multiphoton interference.
In Sec.~\ref{sec:sparse},
we introduce our proposed sparse interferometer and the design principles of the interferometer with the help of graphs.
In Sec.~\ref{sec:measure},
we explicitly examine the case of four photons,
provide a proposal to measure the four-photon collective phase,
and then generalize to the $n$-photon case.
Finally,
we summarize our work in Sec.~\ref{sec:summary}.

\section{Background}
\label{sec:background}
In this section, we summarize key background information and
the state of the art.
In Sec.~\ref{subsec:collectivephase}, we review MCP theory and experiments including genuine multiphoton interference.
In Sec.~\ref{subsec:multiphoton_interference}, we give the basic notation and a theoretical framework for multiphoton interference.
In Sec.~\ref{subsec:graphs}, we introduce the connectivity graph for the interferometer and the enhanced-distinguishability graph for the multiphoton case.

\subsection{Collective phase and genuine multiphoton interference}
\label{subsec:collectivephase}
An MCP was first isolated and observed with a tritter, and this three-photon MCP is known as a ``triad phase''~\cite{Menssen2017Phys.Rev.Lett.}.
The triad phase is defined as the sum of the arguments of the pairwise overlaps of photon states among three photons.
To isolate the triad phase experimentally,
photon polarizations are combined with time delays to control and vary the triad phase so that the overlap amplitude is invariant.
The empirical signature of the triad phase is an oscillating triad phase with respect to varying polarization and time delays.

When one of the three photons is blocked,
the two-photon coincidence is constant under the same conditions that make the triad phase vary.
Constant two-photon coincidence concomitant with varying triad phase is experimental proof of genuine three-photon interference.
Another manifestation of genuine three-photon interference arises for three energy-time entangled photons passing through three Franson interferometers,
each with a distinct, controllable relative phase between the ``short'' and ``long'' paths~\cite{Agne2017Phys.Rev.Lett.}.
The three-photon coincidence rate shows a sinusoidal oscillation with respect to the sum of these three phases.
When one or both photons are traced out, the resultant two-photon coincidence rate or one-photon rate,
respectively,
is invariant with the sum of the phases.

Although both the triad phase~\cite{Menssen2017Phys.Rev.Lett.}
and the three-photon phase~\cite{Agne2017Phys.Rev.Lett.} can show genuine three-photon interference, they have quite different origins.
The triad phase emerges from the relative phase in the internal space of pairwise photons,
whereas the other phase emerges from the Franson interferometer.

Shchesnovich and Bezerra use a weighted directional graph to describe genuine $n$-photon interference with an input corresponding to an $n$-photon separable state arriving and with a scattering matrix that does not have any zero elements~\cite{Shchesnovich2018Phys.Rev.A}.
An interferometer with no zero scattering-matrix elements is called an all-connected interferometer.
In the weighted directed graph representation of multiphoton interference,
each vertex represents a photon, 
and the weight of each directed edge is given by the overlap of the two photons connected by the edge in the internal degree of freedom (DoF).
Then an $n$-photon collective phase corresponds to the cycle of length $n$ in the graph.

For the case of~$n$ photons all mutually interfering in the all-connected interferometer,
the corresponding graph is an $n$-vertex directed complete graph.
Besides the $n$-photon collective phases,
there are also cycles with length less than~$n$;
these smaller cycles correspond to lower-order MCPs in the complete graph.
With graph theory, Shchesnovich and Bezerra prove that genuine $n$-photon interference can be realized by
making all pairs of photons except the neighboring pairs and the first-last pair of photons orthogonal~\cite{Shchesnovich2018Phys.Rev.A}.

By using Shchesnovich and Bezerra's method, 
recent experiments manifest a four-photon collective phase in the four-photon interference of a quitter~\cite{Jones2020Phys.Rev.Lett.}.
In their experiment,
one pair of photons is orthogonal in polarization, 
and the other pair of photons is orthogonal in the time-frequency domain by requiring a large relative-time delay.

Scaling up the experiments,
according to Shchesnovich and Bezerra's method,
to~$n$ photons
requires more independent control parameters over the photons' internal space;
this overhead is clear by counting the number of constraints.
Specifically,
each photon needs to be orthogonal to~$(n-3)$ photons when~$n\geq 3$,
which yields~$n(n-3)$ constraints over the control parameters in the photons' internal DoF space.
If we assume that,
for each photon,
$d$ independent parameters are required to rotate the photon in the internal space,
then
to isolate the $n$-photon collective phase in multiphoton interference,
we require that
\begin{equation}
\label{eq:require}
dn-n(n-3)\geq 0\implies d\geq n-3.
\end{equation}
Thus, the number of independent parameters for controlling each photon increases linearly with respect to the photon number~$n$.

In contrast to all-connected interferometers in Shchesnovich and Bezerra's method,
we double the mode number and then propose a sparse interferometer to manifest genuine $n$-photon interference.
Then we show how to measure the $n$-photon collective phase from the interference results.
We define an interferometer layer as a subset of optical elements that share the same shortest length for input photons to transmit through from the injected ports.
Then,
for our proposed interferometer,
only two layers of optical elements are needed, which gives only constant loss in the interferometer and thereby overcomes the need for independent parameters in the photons' internal DoF space.

\subsection{Multiphoton interference and MCP}
\label{subsec:multiphoton_interference}
In this section,
we introduce basic notation and a known useful expression
for the coincidence rate obtained by multiphoton interference~\cite{Khalid2018Phys.Rev.A,Wu2018Phys.Rev.A}.
First we write the interferometer input state as a product state of~$n$ single photons with the creation operators indexed by external- and internal-DoF labels.
Then we transform these creation operators according to the unitary scattering matrix~$U\in~$SU$(m)$,
where~$m$ is the number of channels.
Finally, we write the expression for the $n$-photon coincidence rate at the output.

Usually,
single photons interfering at an interferometer
involve DoFs arising in two categories.
One category is for DoFs that are transformed by the interferometer,
i.e.,
mixing paths at beam splitters.
The other category is for DoFs that are not manipulated by the interferometer but are rather externally controlled, such as polarization or timing.
The distinction between internal and external DoFs is not always sharp;
for example, doubling the number of interferometer channels to accommodate polarization as additional paths
in effect converts polarization from an internal to an external DoF~\cite{Wu2018Phys.Rev.A}.
These two DoF categories have been called ``system'' and ``label'' DoFs, respectively~\cite{Stanisic2018Phys.Rev.A}.
We instead use the terms ``external'' and ``internal'' DoFs, respectively, for clarity.

We inject~$n$ photons into~$m$ interferometer input ports and label each input photon with a unique index
\begin{equation}
\label{eq:photonindex}
i\in[n]:=\{1,2,\dots,n\}
\end{equation}
and each input port with index~$j\in[m]$.
Then we denote the creation operator for the~$i$th single photon injected into the~$j$th input port as~$A^{(i)\dagger}_j$.
This notation for the creation operator suppresses internal DoFs;
we incorporate internal DoFs by employing the creation operator~$a^\dagger_s$,
where~$s$ labels the internal DoF and the 
external DoF is suppressed.
To be clear,
the expression for the~$i$th single photon entering the~$j$th input port is
\begin{equation}
\label{eq:Aij}
    A^{(i)\dagger}_j\ket0=\sum_sp^{(i)}_{s}a_{j;s}^{\dagger}\ket0,
\end{equation}
with~$\ket0$ being the vacuum state
and
$\bm{p}:=(p_s)$ being a vector of coefficients over the internal DoF.
If~$s$ is treated as being continuous,
such as for time of arrival,
then the sum in Eq.~\eqref{eq:Aij}
is treated instead as an integral.

As the~$n$ input photons are injected into a SU($m$) interferometer,
i.e., a lossless, passive, $m$-channel interferometer,
the interferometer transformation can be described mathematically by an~$m\times m$ unitary scattering matrix
\begin{equation}
\label{eq:unitaryscatteringmatrix}
U=(U_{kj})\in\mathcal{M}_m(\mathbb{C}).
\end{equation}
As we only treat the case that at most one photon is injected into each input port,
we construct an input-configuration vector function
\begin{equation}
\label{eq:v}
\bm{v}:[n]\to[m]:i\mapsto v_i:=\operatorname{proj}_i\bm{v},
\end{equation}
i.e., $v_i$ indicates which input port receives the~$i$th incoming photon.

Given an input port $v_{i}$, the inverse $v^{-1}$ yields the index~\eqref{eq:photonindex}
that pertains to the injected photon,
i.e.,
\begin{equation}
\label{eq:vinverse}
v^{-1}:[m]\to [n]: v_{i}\to i
\end{equation}
Elementwise,
the vector~\eqref{eq:v}
is expressed as
\begin{equation}
\label{eq:vT}
\bm{v}=(v_1 \cdots v_n)^\top,
\end{equation}
with~${}^\top$ denoting transpose.
Note that the order of integers in~$\bm{v}$
is important as permuting which photons enter where can change the multiphoton coincidence measurement outcome.

Now we consider the output configuration.
Although the photons can be regarded as exiting through a superposition of paths,
photon counting of each output path is executed in the end.
Thus we identify an output configuration with postselecting the state based on these photon-count results.
In our analysis,
for output configurations, we postselect over the collision-free case, i.e., at most one photon exits from each output mode.
Thus,
similarly to input-vector~$\bm v$~\eqref{eq:v},
\begin{equation}
\label{eq:eta}
\bm{\eta}:[n]\to[m]:i\mapsto \eta_i,\;
\eta_{i}=\eta_j \iff i=j
\end{equation}
is the output-configuration vector,
with~$\bm\eta$ being a length-$n$ vector of positive integers in the range from 1 to~$m$.
Whereas permuting entries in the input-configuration vector~$\bm{v}$ can change the measurement result,
multiphoton coincidence measurements are set up so that permuting the detectors does not change the results.
Thus, unlike for~$\bm{v}$,
the vector~$\bm{\eta}$ has the same physical meaning under all permutations of its entries.

Now we express the formula for a coincidence rate represented by~$\bm\eta$
and $\bm v$
given input state
\begin{equation}
\label{eq:inputstate}
\Ket{\Psi}_{\text{in}}
    =\prod_{i=1}^{n}A_{v_{i}}^{(i)\dagger}\ket0.
\end{equation}
The output state can be calculated by transforming creation operators by the unitary matrix~$U$ describing the interferometer according to
\begin{align}
\label{eq:AUA}
 A_j^{(i)\dagger}\mapsto
    \sum_{k=1}^{m}U_{k,j}A_{k}^
    {(i)\dagger},\forall i\in[n],j\in[m].
\end{align}
For~S$_n$, the permutation group over~$n$ elements,
we employ 
its regular representation
\begin{equation}
    \left(\Pi_{\sigma}\right)_{i,j}=\begin{cases}
    1&\sigma\sigma_{i}=\sigma_{j}\\
    0&\text{otherwise},
    \end{cases}\quad\sigma_{i},\sigma_{j},\sigma\in\mathrm{S}_{n},
\end{equation}
the interferometer vector
\begin{align}
\left(\bm{u}_{\bm{v}}^{\bm{\eta}}\right)_\sigma:=\prod_{i=1}^{n}U_{\eta_{\sigma(i)},v_{i}},\forall \sigma\in \text{S}_n,
\end{align}
and the permutation-dependent overlaps of photonic states
\begin{equation}
    \label{eq:overlap}
r_\sigma^{(n)}
    =\prod_{i=1}^n\Bra{0}A_{v_{i}}^{\sigma(i)}A_{v_{i}}^{(i)\dagger}\ket0\in\mathbb{C},
    \;\forall\sigma\in\text{S}_n,
\end{equation}
to write the output coincidence rate
as~\cite{Khalid2018Phys.Rev.A,Wu2018Phys.Rev.A}
\begin{equation}
\label{eq:coincidence}
C^{\bm{\eta}}_{\bm{v}}
=\sum_{\sigma\in\text{S}_n}r_\sigma^{(n)}
\left[\left(\bm{u}_{\bm{v}}^{\bm{\eta}}\right)^{\dagger}\cdot \Pi_\sigma\cdot \bm{u}_{\bm{v}}^{\bm{\eta}}\right].
\end{equation}
The coincidence rate is a linear combination of the overlaps $r_{\sigma}^{(n)}$, with each weight given by the expected value of the regular representation $\set{\Pi_{\sigma}}$ 
of the symmetric group S$_n$ with respect to $\bm{u}_{\bm{v}}^{\bm{\eta}}$.

Here, we provide an example of three-photon interference in a tritter~\cite{Zeilinger1993QuantumcontrolandmeasurementElsevier,Campos1989Phys.Rev.A,Menssen2017Phys.Rev.Lett.,Menssen2017Phys.Rev.Lett.} with the scattering matrix defined as 
\begin{equation}
U_{i,j}:=\frac{1}{\sqrt{3}}\mathrm{e}^{-\frac{2\pi\mathrm{i}}{3}(i-1)(j-1)}, \forall i,j\in[3].
\end{equation}
When we inject three photons into the tritter with the input configuration $\bm{v}=(1,2,3)^{\top}$ and detect with the output configuration $\bm\eta=(1,2,3)^{\top}$,
the coincidence rate~\eqref{eq:coincidence} contains six terms,
\begin{equation}
\frac{2}{9}r_{\mathds{1}}^{(3)}-\frac{1}{9}\left(r_{(1,2)}^{(3)}+r_{(2,3)}^{(3)}+r_{(1,3)}^{(3)}\right)+\frac{2}{9}\left(r_{(1,2,3)}^{(3)}+r_{(1,3,2)}^{(3)}\right).
\label{eq:coincidence_three_photon}
\end{equation}
Although the summation of these six terms gives us a real-valued three-photon coincidence rate,
each of them is not necessarily real.

The first term is related to
the overlap of the three-photon state with itself without any permutation,
which gives $r_{\mathds{1}}^{(3)}=1$.
The three terms in the middle of~\eqref{eq:coincidence_three_photon} are related to the two-photon permutations.
We can see that these three terms are real by substituting the two-cycle permutation $\sigma=(j,k)\in\mathrm{S}_{3}$ into Eq.~\eqref{eq:overlap},
\begin{equation}
    r_{(j,k)}^{(3)}=\left|\braket{0|A_{v_k}^{(j)}A_{v_k}^{(k)\dagger}|0}\right|^2,
    \label{eq:rij3}
\end{equation}
where the imaginary phase
is absent due to the modular square.
Each of these three two-photon-related terms can be measured from the Hong-Ou-Mandel interference between the corresponding two photons~\cite{Hong1987Phys.Rev.Lett.}.

However,
the last two terms in~\eqref{eq:coincidence_three_photon} 
can be complex numbers and contain the complex phase that cannot be measured from two-photon interference.
We denote
the phase of the three-photon-interference term $r_{(1,2,3)}^{(3)}$~(or $r_{(1,3,2)}^{(3)}$)
by
\begin{equation}
    \psi_{(1,2,3)}^{(3)}:=\arg \left(r_{(1,2,3)}^{(3)}\right)
\end{equation}
and the phase of the overlap of two photons $j$ and $k$ as
\begin{equation}
    \theta_{j,k}^{(3)}:=\arg\left(\braket{0|A^{(j)}_{v_{k}}A^{(k)\dagger}_{v_{k}}|0}\right),j,k\in[3];
\end{equation}
then,
from Eq.~\eqref{eq:overlap},
we have
\begin{equation}
    \phi_{(1,2,3)}^{(3)}=\theta_{3,1}^{(3)}+\theta_{1,2}^{(3)}+\theta_{2,3}^{(3)}.
\end{equation}
Such an interference phase emergent in the three-photon interference
is a sum of two-photon overlap phases among three photons
and contains the collective information of three photons.
Such a phase is called ``triad phase''~\cite{Menssen2017Phys.Rev.Lett.} or three-photon collective phase due to its analogy with collective behavior in three-photon interference.

Furthermore,
the collective-phase idea can be generalized to more photons.
Although~$C^{\bm{\eta}}_{\bm{v}}$~\eqref{eq:coincidence}
is independent of~$\sigma$ by virtue of the summation over permutations,
the collective phase depends on the overlaps~\eqref{eq:overlap} themselves.
Mathematically,
we decompose the permutation into disjoint product cycles, namely,
\begin{equation}
\sigma=\prod_{i}\sigma_{i}
\implies r_\sigma^{(n)}
=\prod_{i}r_{\sigma_{i}}^{(n)}.
\end{equation}
The~$i$-photon collective phase for an $n$-photon input state~\eqref{eq:inputstate} is
\begin{align}
\psi_\sigma^{(n)}:=\arg (r_\sigma^{(n)}),
    \left|\sigma\right|=i,
\label{eq:photon_phase}
\end{align}
where~$|\sigma|$ denotes the length of a cycle.
Consequently, the two-photon collective phase must be zero because~$\sigma^{-1}=\sigma$ and
\begin{align}
r_{\sigma^{-1}}^{(n)}=r_\sigma^{*(n)},\label{eq:Bsigma}
\end{align}
so~$\sigma$ being a length-2 cycle
implies~$r_\sigma^{(n)}\in\mathbb{R}$.
If~$\sigma$ is a disjoint product of cycles,
then the collective phase is the sum of collective phases for each of these irreducible cycles.
To obtain the whole picture of multiphoton interference, we only need to focus on all the disjoint permutation operators.
\subsection{Connectivity of the interferometer}
\label{subsec:graphs}
For the coincidence rate~\eqref{eq:coincidence}, each term contains a contribution from photonic-state overlaps and the interferometer.
If we make the interferometer sparse, i.e.,
some $U_{i,j}$ elements of the scattering matrix are zero,
the contribution related to some permutations could be diminished.
Our aim is to sparsify the interferometer so that only the contributions related to an $n$-cycle permutation and two-length permutation exist in the coincidence rates.
To better show which terms can exist in a specific interferometer,
in this section,
we employ graphs to represent the interferometer.
From a description of the $m$-channel interferometer by some unitary $U$~(\ref{eq:unitaryscatteringmatrix}),
we construct two graphs to represent this interferometer.
The first graph, the connectivity graph,
represents connectivity,
which reveals whether certain outputs arise given certain inputs.
The second graph,
called the enhanced-distinguishability graph,
represents distinguishability permitted in the interferometer for injected photons.

\begin{figure}
\centering
    (a)~\includegraphics[width=0.23\columnwidth]{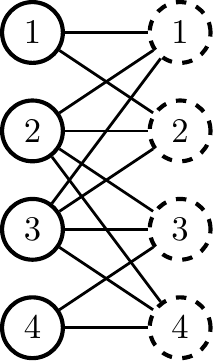}
    (b)~\includegraphics[width=0.27\columnwidth]{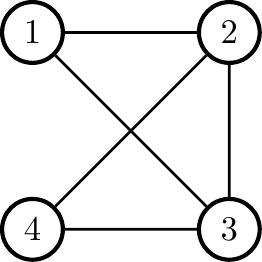}
    (c)~\includegraphics[width=0.27\columnwidth]{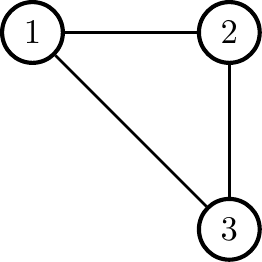}
    \caption{(a)~Connectivity graph of a $4\times 4$ interferometer.
    The input vertices are shown as solid circles, and the output vertices are shown as dashed circles.
    (b)~A minor graph mapped from the connectivity graph in~(a) by the mapping~$L$.
    (c)~The corresponding enhanced-distinguishability graph mapped from (b)~by~$\Theta$ when the three-photon input-configuration vector is~$\bm{v}=(1\,2\,3)^\top$.}
    \label{fig:graph_example}
\end{figure}

The interferometer is described by the unitary matrix explained in Eq.~\eqref{eq:AUA},
which we convert to the connectivity graph,
showing a relation between inputs and outputs,
according to the following rule.
The connectivity graph 
\begin{equation}
\mathsf{G}_{\text{c}}=(\mathsf{E}_{\text{c}},\mathsf{V}_{\text{c}})
\label{eq:Gc}
\end{equation}
for the~$m\times m$ unitary $U$ in Eq.~(\ref{eq:unitaryscatteringmatrix})
has~$2m$ vertices
in \begin{equation}
    \mathsf{V}_{\text{c}}=\mathsf{S}_{\text{c}}\sqcup \mathsf{D}_{\text{c}}
\end{equation}
where $\mathsf{S}_{\text{c}}$ contains
the first~$m$
vertices as solid circles and 
$\mathsf{D}_{\text{c}}$ contains
the last~$m$ vertices as dashed circles,
as seen in Fig.~\ref{fig:graph_example}(a).
For the edge set $\mathsf{V}_{\text{c}}$ of the connectivity graph, if
$U_{ij}\neq0$,
an undirected edge is drawn from the solid circle labeled~$i$
to vertex~$m+j$,
which we draw as a dashed-circle vertex labeled~$j$.
If~$U_{ij}=0$,
no edge exists from the solid circle labeled~$i$ to the dashed circle labeled~$j$.
The connectivity graph has the property that edges only exist between solid and dashed circles,
not from solid circles to solid circles or dashed circles to dashed circles.

Every unitary matrix~$U$ can be mapped to a unique connectivity graph,
but not every connectivity graph can be mapped to a unitary matrix for any choice of~$m$ channels;
that is, the connectivity map is not invertible.
We refer to connectivity graphs that yield unitary matrices as proper connectivity graphs with notation~$\mathfrak{G}_{\text{c}}$,
and the complement are improper connectivity graphs.
In Fig.~\ref{fig:graph_example}(a), we give an example of a proper connectivity graph of a $4\times 4$ interferometer.

Now we explain the mapping from connectivity graphs, 
whether proper or improper,
to enhanced-distinguishability graphs.
In the enhanced-distinguishability graph,
each vertex represents a photon,
and for arbitrary two vertices,
the edge exists only when nonzero distinguishability between the corresponding two photons exists in the coincidence rate~\eqref{eq:coincidence}.
This restriction requires that their two-photon overlap be nonzero and the corresponding two photons interfere in the interferometer.
Here, we assume that the photons are not totally distinguishable,
i.e.,
their distinguishability is nonzero in their internal DoF,
such that we can focus on the relationship between the enhanced-distinguishability graph and the interferometer.

With the above assumption,
to determine whether two photons can interfere in the interferometer,
we can see whether these two photons are intimate in their external DoF,
i.e.,
whether they share at least one common output mode when they exit the interferometer.
Thus,
given the input configuration of photons,
the basic idea of the mapping from the connectivity graph to the enhanced-distinguishability graph is 
to check whether two arbitrary related input vertices in the connectivity graph are connected to at least one common output vertex.

Formally, this mapping is a composition of mappings~$L$ and~$\Theta$,
i.e., $\Theta\circ L$
with
\begin{equation}
\label{eq:L}
L:\mathsf{G}_{\text{c}}
\to\mathsf{G}_{\text{mc}},
\end{equation}
where
$\mathsf{G}_{\text{mc}}=(\mathsf{E}_{\text{mc}},\mathsf{V}_{\text{mc}})$
with the subscript ``mc'' referring to ``minor of connectivity'',
and 
\begin{equation}
\label{eq:Theta}
\Theta:\mathsf{G}_{\text{mc}}
\to\mathsf{G}_{\text{e}}\subseteq\mathsf{G}_{\text{mc}},
\end{equation}
where~$\mathsf{G}_{\text{e}}$ is
the enhanced-distinguishability graph.

The mapping~$L$ is achieved
by lifting all of the solid-dashed-solid paths to solid-solid paths.
This mapping is subdivided into a vertex mapping and an edge mapping as $L=(f_L,g_L)$ with
\begin{equation}
    \label{eq:cgraphtomcgraphvertices}
    f_{L}: \mathsf{V}_{\text{c}}\to \mathsf{V}_{\text{mc}}: x\mapsto \begin{cases}
    x & \quad\text{if } x\in \mathsf{S}_{\text{c}}\\
    \emptyset & \quad \text{if } x\in \mathsf{D}_{\text{c}}
    \end{cases}
\end{equation}
for vertices
and,
for an edge defined by its pair of vertices $(x,y)$,
a mapping
\begin{widetext}
\begin{equation}
    \label{eq:cgraphtomcgraphedges}
    g_{L}:\mathsf{E}_{\text{c}}\to \mathsf{E}_{\text{mc}}:
    (x,y)
    \mapsto \begin{cases}
    (x,z) &\text{if } x,z\in\mathsf{S}_{\text{c}} \land y\in \mathsf{D}_{\text{c}}\land (y,z)\in \mathsf{E}_{\text{c}}\\
    \emptyset &\text{otherwise}
    \end{cases}
\end{equation}
for the edges, which essentially maps the path $(x,y,z)$ in $\mathsf{G}_{\text{c}}$ to the edge $(x,z)$ in $\mathsf{G}_{\text{mc}}$.
We illustrate the mapping~$L$ in Fig.~\ref{fig:graph_example} for a proper connectivity graph~[Fig.~\ref{fig:graph_example}(a)] mapped to its minor~[Fig.~\ref{fig:graph_example}(b)].
In the example, there is an edge between vertices 1 and 2 in Fig.~\ref{fig:graph_example}(b) because there exists a path $\left(\circled{1},\dashcircled{1},\circled{2}\right)$ in Fig.~\ref{fig:graph_example}(a).
However, we can see that there is no edge between vertices 1 and 4 in Fig.~\ref{fig:graph_example}(b) due to the absence of a solid-dashed-solid path that connects $\circled{1}$ and $\circled{4}$ in Fig.~\ref{fig:graph_example}(a).

From Eq.~\eqref{eq:cgraphtomcgraphedges},
we see that there is a mapping from the edge of a solid-dashed-solid path to the dashed vertices in the path, which is denoted by 
\begin{equation}
\label{eq:pathtodash}
    h:\mathsf{E}_{\text{c}}\to \mathsf{D}_{\text{c}}:
    (x,y)\mapsto
    \begin{cases}
    y&\text{if }
    x\in\mathsf{S}_{\text{c}}
    \land
    y\in\mathsf{D}_{\text{c}}
    \land
    \exists z\in\mathsf{S}_{\text{c}}
    \land 
    (y,z)\in \mathsf{E}_{\text{c}},\\
    \emptyset &\text{otherwise.}
    \end{cases}
\end{equation}
\end{widetext}
The mapping~$h$ and its inverse are used in Sec.~\ref{subsec:doubling}, where we explain the reason for doubling the mode number in the interferometer to achieve genuine multiphoton interference.

Next, when we specify the input configuration of photons as~$\bm{v}$~\eqref{eq:v},
we only care about those vertices labeled by each of the components in~$\bm{v}$.
This gives us the mapping $\Theta$ that takes a $\mathsf{G}_{\text{mc}}$ to a $\mathsf{G}_{\text{e}}$,
which is constructed by picking up vertices labeled by~$\bm{v}$ and the edges that are linked to the vertices picked in $\mathsf{G}_{\text{mc}}$. 
Similarly, given~$\bm{v}$ and~$\mathsf{G}_{\text{mc}}=(\mathsf{E}_{\text{mc}},\mathsf{V}_{\text{mc}})$, 
the mapping~$\Theta=(f_\Theta,g_\Theta)$ comprises
\begin{equation}
\label{eq:cgraphtoedgraphvertices}
f_{\Theta}:\mathsf{V}_\text{mc}
\to\mathsf{V}_\text{e}:
x\mapsto \begin{cases}
v^{-1}(x)&\text{if } x\in\left\{\operatorname{proj}_i\bm{v}\right\}\\
\emptyset &\text{otherwise}
\end{cases}
\end{equation}
for the vertex mapping, with~$v^{-1}$ defined in Eq.~\eqref{eq:vinverse},
and
\begin{align}
\label{eq:cgraphtoedgraphedges}
g_{\Theta}:&\,\mathsf{E}_\text{mc}\to\mathsf{E}_\text{e}:\nonumber\\
&(x,y)\mapsto\begin{cases}
\left(v^{-1}(x),v^{-1}(y)\right) & \text{if }x,y\in\left\{\operatorname{proj}_i\bm{v}\right\}\\
\emptyset&\text{otherwise}
\end{cases}
\end{align}
for the edge mapping.

Our first step in this mapping~$\Theta$ is to pick up vertices corresponding to labels for input ports receiving single photons
from the minor~$\mathsf{G}_{\text{mc}}$.
The remaining vertices that are not included in the enhanced-distinguishability graph
represent labels for ports receiving only vacuum.
We relabel these picked vertices with labels from 1 to~$n$ following the inverse mapping $v^{-1}$ in~\eqref{eq:vinverse}.
Next we draw an edge connecting vertices if the corresponding two vertices are adjacent in~$\mathsf{G}_{\text{mc}}$.
We show this mapping in the example in Fig.~\ref{fig:graph_example} from the minor of the connectivity graph~[Fig.~\ref{fig:graph_example}(b)] to the enhanced-distinguishability graph~[Fig.~\ref{fig:graph_example}(c)] if the vector~\eqref{eq:vT}
is
$\bm{v}=(1\,2\,3)^{\top}$.
A special case arises for~$\bm{v}=(1\,2\,\cdots\,n)^{\top}$
and~$n=m$:
In this case, $\mathsf{G_{\text{mc}}}$ coincides with~$\mathsf{G}_{\text{e}}$.

Our formalism for connectivity graphs and enhanced-distinguishability graphs applies to the special case of all-connected interferometers~\cite{Shchesnovich2018Phys.Rev.A},
for which a single photon entering any of the input ports has a nonzero probability of being detected at any output port.
Consequently,
the interferometer's connectivity graph is a complete bipartite graph between solid-circle and dashed-circle vertices;
i.e., each input vertex is connected to all output vertices. 
This all-connected graph yields an enhanced-distinguishability graph
that is necessarily a complete graph,
i.e., all vertices are mutually connected.

\section{Sparse interferometer}
\label{sec:sparse}
To realize genuine $n$-photon interference and measure the $n$-photon collective phase with separable photons,
the basic idea is to eliminate all the lower-order phases from the coincidence.
In Sec.~\ref{subsec:concept},
we elucidate our concept for sparse interferometers eliminating lower-order MCPs.
Then in Sec.~\ref{subsec:doubling}, we analyze the necessity of doubling the modes of the interferometer to generate the genuine $n$-photon interference and isolate the $n$-photon collective phase in our method.
In Sec.~\ref{subsec:configurations}, 
we sort the output configurations into three sets such that one set is only related to these pairwise overlaps of photons, 
one set is only related to the $n$-photon collective phase, 
and the remaining set contains all forbidden configurations in our setup.

\subsection{Concept}
\label{subsec:concept}
Currently, only one method exists for eliminating all lower-order MCPs.
This method involves manipulating all photons' internal DoFs
so that all lower-order phase terms in the coincidence disappear due to total distinguishability between selected photon pairs~\cite{Shchesnovich2018Phys.Rev.A,Jones2020Phys.Rev.Lett.}.
We develop a method based on manipulating external DoFs, i.e.,
using a sparse interferometer.
In our sparse interferometer the number of nonzero elements in the corresponding~$m\times m$ scattering matrix is~$O(m)$,
unlike $m^2$ in an all-connected interferometer.

Here we design a sparse~$2n$-path interferometer whose output photon configuration,
given a valid $n$-photon input configuration
(either one or zero photons at each input port),
depends only on the~$n$th MCP and not on any MCP with order less than~$n$.
Furthermore,
we require that the longest optical path in the sparse interferometer is constant regardless of~$n$.
Our interferometer-design algorithm accepts the $n$-photon input configuration and the enhanced-distinguishability graph;
the algorithmic output is a family of~$2n$-path interferometer designs whose outputs all depend only on the the~$n$th MCP as required.

\subsection{Doubling the modes}
\label{subsec:doubling}
Our design procedure commences somewhat paradoxically with the requirement that we double the number of modes to achieve an efficacious sparse interferometer.
As we seek to reduce optical depth,
increasing the number of modes seems strange,
but this approach is effective because only a constant overhead in optical depth is required if modes are doubled.
However, increasing the number of modes is more than compensated by the ability to sparsify the interferometer as we show in this section.

\begin{figure}
\centering
    (a)~\includegraphics[width=0.24\columnwidth]{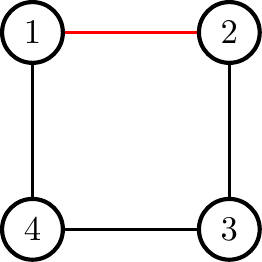}
    (b)~\includegraphics[width=0.21\columnwidth]{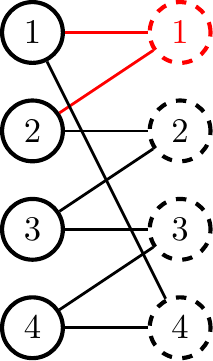}
    (c)~\includegraphics[width=0.34\columnwidth]{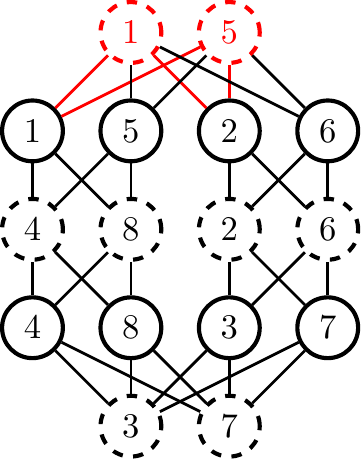}
    \caption{%
    (a)~Enhanced-distinguishability graph with only a cycle of length 4, i.e., $\sigma=(1,2,3,4)$. When the input-configuration vector $\bm{v}=(1\,2\,3\,4)^{\top}$, 
    the corresponding connectivity graphs with four input vertices~(shown as solid circles) 
    and four output vertices~(shown as dashed circles) in~(b) and eight input vertices and eight output vertices in~(c). 
    The edge between input vertices~1 and 2 is highlighted in red in (a), 
    and the corresponding~$\left.O_{1,2}^{(4)}\right|_{\mathsf{G}_{\text{c}}}$ in connectivity graphs (b) and~(c) are also in red.}
    \label{fig:connect_4_4}
\end{figure}

Our approach is to design a sparse interferometer from the input~$n$ and an enhanced-distinguishability graph that must be topologically a cycle of length~$n$.
Specifically,
we design an interferometer with~$n$ photons in input configuration~$\bm{v}$~\eqref{eq:v}, requiring an enhanced-distinguishability graph that is a cycle of length~$n$,
for example, the cycle~$(1,2,3,4)$ with four ($n=4$) vertices in Fig.~\ref{fig:connect_4_4}(a).

To reach our goal, we first explain why we need our input configuration to be in the form of Eq.~\eqref{eq:v}, where there is at most one photon occupied at each input port.
To explain this restriction,
we begin by considering the forbidden case of more than one photon entering an input port,
and then show that more than one photon entering any input port
leads to cycles of length less than~$n$ existing in the enhanced-distinguishability graph,
implying the influence of lower-order MCPs, which we are trying to avoid.

Let us make this argument clear and explicit.
Suppose two photons~$i$ and~$j$ are injected into the same input port.
The corresponding vertices~$i$ and~$j$ are consequently connected together within the enhanced-distinguishability graph.
As each vertex is connected to two other vertices in a cycle,
a third vertex~$k$ other than~$i$ and~$j$ is connected to~$i$, as well as to~$j$.
Thus these three vertices are connected to each other in the enhanced-distinguishability graph,
which yields a cycle of length~3 and consequently the undesired signature of the triad phase cluttering the higher-order MCP measurement.
Furthermore, when more than two photons are injected into the same input port,
it is easy to see that the corresponding vertices are mutually connected in the enhanced-distinguishability graph,
which gives us cycles of length less than~$n$.

Without loss of generality, we express the input configuration as~$\bm{v}=(1\,2 \cdots\,n)^\top$, 
and the enhanced-distinguishability graph is an $n$-length cycle graph~$(1,2,3,\dots,n)$.
The enhanced-distinguishability graph can be retrieved from the connectivity graph by the mapping~$L^{-1}\circ\Theta^{-1}$,
which is not necessarily unique
because the connectivity graph could 
have more than one solid-dashed-solid path connecting two solid vertices.
To consume fewer optical elements,
what we want is to compute the minimum number of solid-dashed-solid paths required for the connectivity graph to be proper,
i.e., in~$\mathfrak{G}_{\text{c}}$.

This minimum also informs us regarding the minimum number of modes required for observing the MCP without lower-order terms because of the mapping~$h$~\eqref{eq:pathtodash} from the edges of solid-dashed-solid paths to the dashed vertices in those paths.
We denote a graph of the $n$-length cycle by~$\mathsf{C}^{(n)}$.
The enhanced-distinguishability graph of interest occurs for $\mathsf{G}_{\text{e}}=\mathsf{C}^{(n)}$.
Then we denote an arbitrary edge from the enhanced-distinguishability graph of interest as~$(i,j)$.

Given a connectivity graph $\mathsf{G}_{\text{c}}$~\eqref{eq:Gc},
we define
\begin{equation}
\label{eq:hgL-1gTheta-1}
\left.O_{i,j}^{(n)}\right|_{\mathsf{G}_{\text{c}}}
:=\Set{x\in \mathsf{D}_{\text{c}}|g_{\Theta}\circ g_{L}\circ h^{-1}(x)=(i,j)},
\end{equation}
comprising all dashed vertices in the connectivity graph that can be mapped to~$(i,j)$ by~$g_{\Theta}\circ g_{L}\circ h^{-1}$,
i.e.,
whose cardinality
$\left|\left.O_{i,j}^{(n)}\right|_{\mathsf{G}_{\text{c}}}\right|$
is the number of dashed vertices.
Then the problem becomes to compute the minimum number of dashed vertices inside each~$\left.O_{i,j}^{(n)}\right|_{\mathsf{G}_{\text{c}}}$
over different choices of the connectivity graph~$\mathsf{G}_{\text{c}}$, 
with this minimum not necessarily being unique.
This minimum is the subset
\begin{align}
\label{eq:minO}
\mathfrak{G}_{\text{c}}^\text{min}
:=\operatorname*{arg\,min}_{\mathsf{G}_{\text{c}}\in\mathfrak{G}_{\text{c}}}\left|\left.O_{i,j}^{(n)}\right|_{\mathsf{G}_{\text{c}}}:
\Theta \circ L:\mathsf{G}_{\text{c}}\mapsto\mathsf{C}^{(n)}\right|
\subset\mathfrak{G}_\text{c},
\end{align}
where $(i,j)$ can be any edge chosen from $\mathsf{C}^{(n)}$.
Now we proceed to explain how to find a~$\mathsf{G}_{\text{c}}\in\mathfrak{G}_{\text{c}}^{\min}$.

Without loss of clarity, we suppress~$\mathsf{G}_{\text{c}}$ in~$\left.O_{i,j}^{(n)}\right|_{\mathsf{G}_{\text{c}}}$ in the following.
When we fix the $n$-cycle graph in~$\mathsf{C}^{(n)}$ to be $\sigma=(1,2,3\dots,n)$,
we can sort all the dashed vertices into~$n$ sets
\begin{equation}
\label{eq:nsets}
O_{i,i+1}^{(n)};
    i\in[n]\;
\left(O_{n,n+1}^{(n)}\equiv O_{n,1}^{(n)}\right).
\end{equation}
There is no intersection between any two distinct sets~\eqref{eq:nsets};
that is,
\begin{equation}
O_{i,i+1}^{(n)}\cap O_{j,j+1}^{(n)}=\emptyset\,
\forall i\neq j\in[n].
\end{equation}
This intersection is empty;
otherwise, there would exist at least one dashed vertex that is connected to more than two solid vertices in the connectivity graph, which leads to the existence of the triad phase among them,
hence a contradiction. 
Examples of~$n=4$ are given in Fig.~\ref{fig:connect_4_4}, 
where the edge in red in Fig.~\ref{fig:connect_4_4}(a) corresponds to the dashed vertices in red in Figs.~\ref{fig:connect_4_4}(b) and~\ref{fig:connect_4_4}(c) 
by the mapping in~\eqref{eq:hgL-1gTheta-1} with~$\bm{v}=(1\,2\,3\,4)$. 
We see that~$O_{1,2}^{(4)}=\set{1}$ in Fig.~\ref{fig:connect_4_4}(b) and~$O_{1,2}^{(4)}=\set{1,5}$ in Fig.~\ref{fig:connect_4_4}(c).

Now we show that the minimum number of dashed vertices in each set is 2 for the constraint optimization problem~\eqref{eq:minO}. 
Suppose there is one dashed vertex in each set;
then there needs to be~$m=n$ output modes,
such as in an example of~$n=4$ shown in Fig.~\ref{fig:connect_4_4}(b).
The corresponding scattering matrix can be expressed as
\begin{align}
    \begin{pmatrix}
    w_{1,1}& w_{1,2}&0&0&\cdots&0&0\\
    0&w_{2,2}&w_{2,3}&0&\cdots&0&0\\
    \vdots&\vdots&\vdots&\vdots&\ddots&\vdots&\vdots\\
    0&0&0&0&\cdots&w_{n-1,n-1}&w_{n-1,n}\\
    w_{n,1}&0&0&0&\cdots&0&w_{n,n}
    \end{pmatrix}.\label{eq:nonunitary}
\end{align}
As the inner product between arbitrary pairs of neighboring columns or rows in~\eqref{eq:nonunitary} is nonzero,
this matrix cannot be unitary.
To overcome the problem so that the corresponding scattering matrix can be unitary to correspond to a lossless linear interferometer~\cite{Yurke1986Phys.Rev.A,Campos1989Phys.Rev.A},
we can add one more dashed vertex to the set.

If we add one more dashed vertex labeled~$(n+i)$ in~$O_{i,i+1}^{(n)}$,
then the~$i$th and~$(i+1)$th columns become
\begin{align}
\begin{pmatrix}
\cdots&w_{i-1,i}&w_{i,i}&0&\cdots &w_{n+i,i}&\cdots\\
\cdots&0&w_{i,i+1}&w_{i+1,i+1}&\cdots &w_{n+i,i+1}&\cdots
\end{pmatrix}^{\top}.
\end{align}
We restrict the inner product between the~$i$th and~$(i+1)$th columns 
by setting
\begin{equation}
w_{i+1,i}^{*}w_{i+1,i+1}+w_{i+1,i}^{*}w_{n+i,i+1}=0,
\end{equation}
which makes these two columns orthogonal.

Following the same procedure,
the nonorthogonality problem of the rest of the~$n-1$ neighboring pairs of columns can be solved by adding one more output vertex for each of the~$n-1$ input neighboring pairs. 
Similarly, the neighboring pairs of rows become orthogonal by adding one more solid vertex~$(n+i)$ to connect the dashed vertices~$(i-1)\bmod n$ and~$i$, 
and the newly added dashed vertices~$[n+(i-1)\bmod n]$ and~$(n+i)$.
Then the resultant connectivity graph is proper as in the example of~$n=4$ in Fig.~\ref{fig:connect_4_4}(c).
Thus we conclude that we need at least~$m=2n$ modes to construct a lossless passive linear interferometer whose corresponding enhanced-distinguishability graph is an $n$-cycle graph if~$\bm{v}=(1\,2\,\cdots\,n)^{\top}$.

To recap,
with the help of the connectivity graph and enhanced-distinguishability graph,
we have shown that,
with a matrix in which each row and column contains only two nonzero elements,
we can realize $n$-photon interference
with contributions only from three classes of permutations of photons: the identity, the two-cycle permutations and an $n$-cycle permutation that corresponds to the $n$-photon collective phase.
However,
such a matrix cannot be unitary as indicated from Eq.~\eqref{eq:nonunitary}.
Thus
we double the mode number and embed the matrix in a bigger unitary matrix
in which each row has four nonzero elements, two of which are in common columns with two elements in the nearest row.
This doubling procedure 
allows us to retain only a
single $n$-photon phase in a $2n\times 2n$ unitary matrix.

\subsection{Output configurations}
\label{subsec:configurations}
Thus far,
we have given the rules to design a sparse interferometer where there is only an $n$-photon collective phase and there are no lower-order MCPs manifested in the interference phenomena.
However, not all output-configuration multiphoton coincidences,
labeled by output-configuration vectors $\{\bm\eta\}$~\eqref{eq:eta},
reveal the $n$-photon collective phase,
so choosing an appropriate subset of configurations~$\{\bm\eta\}$ is important.
One subset of output configurations comprises output configurations that manifest the $n$-photon collective phase,
another subset comprises output configurations that are only related to pairwise overlaps,
and the final subset comprises the rest of the output configurations that are forbidden by our scheme.

In this section, we sort all these $n$-photon output configurations into these three subsets with the help of the connectivity graph and enhanced-distinguishability graph.
Note that the $n$-cycle graph in the enhanced-distinguishability graph need not be~$(1,2,\dots,n)$, 
but could be a cycle permuted by an arbitrary~$\rho\in \text{S}_n$.
Thus, in this section, 
we regard
\begin{equation}
\label{eq:ncyclerho}
\sigma=\left(\rho(1),\rho(2),\cdots,\rho(n)\right)
\end{equation}
as the $n$-cycle graph in the enhanced-distinguishability graph.

For the design of the connectivity graph in Sec.~\ref{subsec:doubling},
we take all the output modes into consideration.
In contrast,
this section,
we clarify which output configurations are related to the $n$-photon collective phase and which are only related to these pairwise overlaps.
Thus, when we specify an output configuration~$\bm{\eta}$,
we can construct a subgraph of the original connectivity graph by deleting the dashed vertices whose labels are not in~$\bm{\eta}$.
Then with the input configuration~$\bm{v}$,
this subgraph can be further mapped to an enhanced-distinguishability graph by~$\Theta\circ L$ as described in Sec.~\ref{subsec:graphs}.

If the enhanced-distinguishability graph for an output configuration is a cycle of length~$n$,
we put the corresponding output configuration represented by length-$n$ integer-valued vectors~\eqref{eq:eta}
in the subset
\begin{align}
\label{eq:xin}
    \xi^{(n)}:=\Set{\bm{\eta}| \forall i\in [n],\,\eta_{i}\in O_{\rho(i),\rho(i+1)}^{(n)}},
\end{align}
which comprises all output configurations whose multiphoton coincidences are related to the $n$-photon collective phase.
Each element in this set is constructed by picking up one output vertex in each set~$O_{\rho(i),\rho(i+1)}^{(n)}$ with~$i\in[n]$.

If there exists an isolated vertex in the enhanced-distinguishability graph,
we put the corresponding output configurations in the subset
$\zeta^{(n)}\subset\{\bm\eta\}$, 
which contains all output configurations forbidden in our setup.
For an output configuration~$\bm{\eta}$, 
if there is an isolated vertex in the enhanced-distinguishability graph,
a solid vertex is not connected to any dashed vertex in the corresponding subgraph of the connectivity graph,
which means that the corresponding photon either is lost or exits from an output mode other than the output modes in~$\bm{\eta}$.
The former
(photon loss)
is impossible for our unitary interferometer. 
The latter
leads to zero coincidence for the output configuration~$\bm{\eta}$.
Thus we conclude that
\begin{align}
    \zeta^{(n)}:=\Set{\bm{\eta}|
    \begin{aligned}
        &\exists\,i\in[n]\, \forall j\in[n]:\\
        &\eta_{j}\notin O_{\rho(i),\rho(i+ 1)}^{(n)}\sqcup O_{\rho(i),\rho(i- 1)}^{(n)}
    \end{aligned}
        }
\end{align}
comprises all forbidden output configurations in our settings for the interferometer and the injected photons.

The last subset
$\chi^{(n)}\subset\{\bm\eta\}$
comprises the output configurations whose multiphoton coincidences are related to the two-photon overlaps but not with the $n$-photon collective phase.
This set can be constructed by making the complement of the union of~$\xi^{(n)}$ and~$\zeta^{(n)}$,
i.e.,
\begin{align}
    \label{eq:chin}
    \chi^{(n)}
    :=\Set{\bm{\eta}|\forall i\in[n],\eta_{i}\in[2n]}
    \setminus\xi^{(n)}\sqcup \zeta^{(n)}.
\end{align}
For the output configurations in~$\chi^{(n)}$, 
we remove the output configurations that are related to two or more pairwise overlaps.
Also we sort the remaining output configurations into~$(n-1)$ subsets.
Each subset~$\chi^{(n)}_{\rho(i),\rho(i+1)}$ contains the output configurations that are only related to the pairwise overlap
\begin{equation}
\label{eq:pairwiseoverlap}
r_{(\rho(i),\rho(i+1))}^{(n)}.
\end{equation}

As most detectors are not photon-number-resolving detectors, 
we only consider collision-free cases, i.e., 
there is at most one photon in each output mode.
Then each subset~$\chi^{(n)}_{\rho(i),\rho(i+1)}$ can be constructed by choosing two vertices in~$O^{(n)}_{\rho(i),\rho(i+1)}$
and then picking up the~$(n-2)$ vertices from the rest of the~$(n-1)$~$O^{(n)}$ subsets with at most one vertex picked up from each set. 
Formally,
\begin{equation}
\label{eq:chinrhoirhoi+1}
    \chi_{\rho(i),\rho(i+1)}^{(n)}:= \Set{\bm{\eta}|
    \begin{aligned}
    &\exists j,k\in[n],\eta_{j}\neq\eta_{k}\in  O^{(n)}_{\rho(i),\rho(i+1)};\\
    &\exists S\subset [n]\setminus\set{i},|S|=n-2,\\
    &\forall j\in S,\exists l\in [n],\eta_{l}\in O_{\rho(j),\rho(j+1)}^{(n)}
    \end{aligned}
    }
\end{equation}
for all~$i\in[n]$.

Multiphoton coincidences corresponding to output configurations in~$\chi^{(n)}_{\rho(i),\rho(i+1)}$ can be used to extract the pairwise overlaps~\eqref{eq:pairwiseoverlap}.
With this pairwise-overlap information,
we can further extract the $n$-photon collective phase from multiphoton coincidences of output configurations in~$\xi^{(n)}$.
In the next section, we give an explicit construction of the interferometer and a concrete example of measuring the MCP.

\section{Measuring the collective phase}
\label{sec:measure}
In Sec.~\ref{sec:sparse}, 
we have provided the interferometer design 
such that each input port only connects to two output ports with the requirement that an $n$-photon collective phase is the only MCP revealed by the interference phenomenon.
In this section, we explicitly describe the interferometer setup
and provide our proposal for measuring the MCP with the interferometer.
In Sec.~\ref{subsec:four-photon}, we present our example for the four-photon case.
We generalize to give an explicit construction for the $n$-photon case in Sec.~\ref{subsec:multiphoton}
for arbitrary~$n$.
\subsection{Four photons and eight-mode interferometer}\label{subsec:four-photon}
\begin{figure}
\centering
    (a)~\includegraphics[width=0.3\columnwidth]{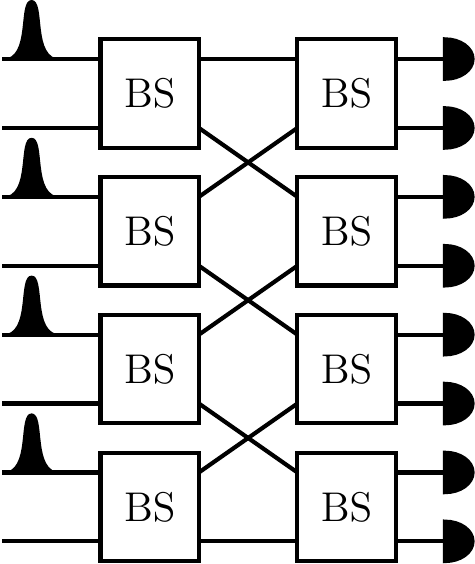}
    (b)~\includegraphics[width=0.3\columnwidth]{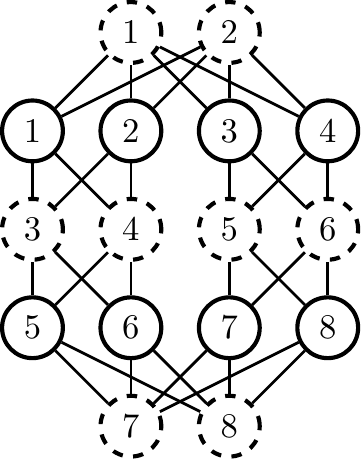}
    (c)~\includegraphics[width=0.2\columnwidth]{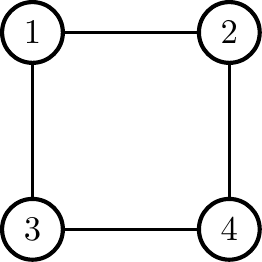}
    \caption{(a)~Sparse interferometer consisting of two layers of BSs for genuine four-photon interference.
    Four 50:50 BSs are placed in each layer.
    The four photons are injected with input-configuration vector
    $\bm{v}=(1\,3\,5\,7)^{\top}$.
    (b)~The corresponding connectivity graph. (c)~The corresponding enhanced-distinguishability graph.
    }
    \label{fig:interferometerfour}
\end{figure}

In this section,
we explicitly construct the~$8\times 8$ interferometer for four-photon interference as shown in Fig.~\ref{fig:interferometerfour}(a),
where there are two layers of 50:50 BSs with four BSs 
labeled by 1, 2, 3, and 4 from top to bottom in each layer.
Each 50:50 BS in Fig.~\ref{fig:interferometerfour}(a) is described by the scattering matrix
\begin{equation}
\frac{1}{\sqrt{2}}\begin{pmatrix}
    1&-1\\
    1&1
\end{pmatrix},
\end{equation}
and the labels for the input or output ports are~1 and~2 from top to bottom.

In our setup, each photon is injected at input port~1 of each BS in the first layer.
Then the photon transmitting through BS 2~(or 3) in the first layer 
has the chance to transmit to two BSs, BSs 1 and 3~(or 2 and 4).
Also,
the photon transmitting from BS 1~(or 4) in the first layer
has the chance to transmit to BSs 1 and 2~(or 3 and 4) in the second layer.
Additionally, a single-photon detector is placed at each output port of the interferometer to record multiphoton coincidence events.

To measure the four-photon MCP experimentally,
we repeatedly inject four photons from input ports 1, 3, 5, and 7, respectively,
and count all four-photon coincidence events.
We calculate the coincidence rates for 
the output configurations in
\begin{equation}
\label{eq:chi4}
\chi^{(4)}_{1,2},\,\chi^{(4)}_{2,4},\,\chi^{(4)}_{4,3},\,\chi^{(4)}_{3,1}
\end{equation} 
and in $\xi^{(4)}$.
Then the four pairwise overlaps 
\begin{equation}
\label{eq:r4}
    r_{1,2}^{(4)},\,r_{2,4}^{(4)},\,r_{4,3}^{(4)},\,r_{3,1}^{(4)}
\end{equation} 
can be calculated from the coincidence rates of the output configurations in the four subsets in Eq.~\eqref{eq:chi4}.
Using the four pairwise overlaps~\eqref{eq:r4},
we retrieve the four-photon MCP from the coincidence rates for output configurations in $\xi^{(4)}$ with a formula that we present in this section.
In the following, we derive these formulas in detail.

With the four-photon input configuration
\begin{equation}
\label{eq:v1357}
\bm{v}=(1\,3\,5\,7)^{\top},
\end{equation}
we show the corresponding connectivity graph in Fig.~\ref{fig:interferometerfour}(b).
We thus obtain the enhanced-distinguishability graph in Fig.~\ref{fig:interferometerfour}(c), which yields an enhanced-distinguishability graph of a cycle~$(1,2,4,3)$, which is permuted from~$(1,2,3,4)$ by~$\rho=(3,4)$. Additionally,
\begin{align}
&O^{(4)}_{1,2}=\set{1,2},\,
O^{(4)}_{2,4}=\set{5,6},\nonumber\\
&O^{(4)}_{3,4}=\set{7,8},\,
O^{(4)}_{1,3}=\set{3,4}
\end{align}
gives us all the~$O^{(4)}$ sets in the example of Fig.~\ref{fig:interferometerfour}.

To calculate the four-photon coincidence rate,
we write the SU(8) scattering matrix
\begin{equation}
    U=\frac12\begin{pmatrix}
        1 & -1 & -1 & 1 & 0 & 0 & 0 & 0\\ 
        1 & -1 & 1 & -1 & 0 & 0 & 0 & 0\\ 
        1 & 1 & 0 & 0 & -1 & 1 & 0 & 0\\ 
        1 & 1 & 0 & 0 & 1 & -1 & 0 & 0\\ 
        0 & 0 & 1 & 1 & 0 & 0 & -1 & 1\\ 
        0 & 0 & 1 & 1 & 0 & 0 & 1 & -1\\
        0 & 0 & 0 & 0 & 1 & 1 &-1 & -1 \\ 
        0 & 0 & 0 & 0 & 1 & 1 & 1 & 1
        \end{pmatrix}
\end{equation}
for the interferometer.
For arbitrary~$i,j\in[4]$, each~$O_{i,j}^{(4)}$ comprises at most two dashed vertices in the connectivity graph of this interferometer.
Thus, when we focus on any~$\bm{\eta}\in\chi^{(4)}$,
there is at least one set~$O_{\rho(i),\rho(i+1)}^{(4)}$
such that there exists two elements of~$\bm{\eta}$, say, $\eta_{s}$ and~$\eta_{t}$ with~$s,t\in [4]$, 
in~$O_{\rho(i),\rho(i+1)}^{(4)}$.
In other words,
\begin{equation}
\label{eq:etaetatinO4}
\eta_{s},\eta_{t}\in O_{\rho(i),\rho(i+1)}^{(4)}.
\end{equation}
All such~$O_{\rho(i),\rho(i+1)}^{(4)}$ give us terms related to the corresponding pairwise overlaps
\begin{equation}
\label{eq:pairwiseoverlaps4}
\left\{r_{(\rho(i),\rho(i+1))}^{(4)}\right\}
\end{equation}
in the multiphoton-interference coincidence rate.
For example, when we choose the output configuration 
$\bm{\eta}=(1\,2\,7\,8)^{\top}$, 
we note that both~$O_{1,2}^{(4)}$ and~$O_{3,4}^{(4)}$ are such related output vertex sets.
Then we see that both~$r_{(1,2)}^{(4)}$ and~$r_{(3,4)}^{(4)}$ are in the coincidence-rate expression
\begin{equation}
\label{eq:Cv1278T}
    C^{(1\,2\,7\,8)^{\top}}_{\bm{v}}=\frac1{2^{7}}\left[2-r_{(1,2)}^{(4)}-r_{(3,4)}^{(4)}+r_{(1,2)}^{(4)}r_{(3,4)}^{(4)}\right]
\end{equation}
calculated from Eq.~\eqref{eq:coincidence}.

The two pairwise overlaps
$r_{(1,2)}^{(4)}$ 
and~$r_{(3,4)}^{(4)}$~\eqref{eq:pairwiseoverlaps4}
are mapped to a particular coincidence-rate expression $C^{(1\,2\,7\,8)^{\top}}_{\bm{v}}$
according to Eq.~\eqref{eq:Cv1278T},
but inverting this coincidence-rate expression to obtain the two pairwise overlaps
is not unique and thus involves constraints to find the correct pair of solutions to this inversion problem.
Here, we focus on output configurations in~\eqref{eq:chi4}.
An example is
\begin{equation}
\bm{\eta}=(1\,2\,3\,5)^{\top}\in\chi^{(4)}_{1,2}:
\eta_{1},\eta_{2}\in O^{(4)}_{1,2},
\end{equation}
and the coincidence-rate expression is
\begin{align}
    C^{(1\,2\,3\,5)^{\top}}_{\bm{v}}
    =\frac1{2^{7}}\left(1-r_{(1,2)}^{(4)}\right).
\end{align}
There are 11 more output configurations in~$\chi^{(4)}_{1,2}$ 
that give the same multiphoton coincidence rate.
We sum these rates to obtain
\begin{align}
    \sum_{\bm{\eta}\in\chi^{(4)}_{1,2}}C^{\bm{\eta}}_{\bm{v}}=\frac{3}{2^{5}}\left(1-r_{(1,2)}^{(4)}\right).
\end{align}
Then the pairwise overlap~$r_{(1,2)}^{(4)}$ is retrieved from the coincidence rates of output configurations in~$\chi^{(4)}_{1,2}$.
Following the same procedure, we obtain
all pairwise overlaps in Eq.~\eqref{eq:r4} as
\begin{equation}
\label{eq:rrhoirhoi+1}
   r_{(\rho(i),\rho(i+1))}^{(4)}=1-\frac{2^{5}}{3}\sum_{\bm{\eta}\in\chi_{\rho(i),\rho(i+1)}^{(4)}}C_{\bm{v}}^{\bm{\eta}},\,
   i\in[4],
\end{equation}
with
$r^{(4)}_{\rho(4),\rho(5)}\equiv r^{(4)}_{\rho(4),\rho(1)}$.

For~$\bm{\eta}\in \xi^{(4)}$, if we define the parity of the output configurations as
\begin{equation}
\operatorname{par}(\bm{\eta})
:=\prod_{i=1}^{n}(-1)^{\eta_{i}},
\end{equation}
with~$\operatorname{par}(\bm{\eta})=1$~($-1$) being the even~(odd) parity,
then output configurations with different parities give different coincidence rates.
For example, $\bm{\eta}=(1\,3\,5\,7)^{\top}$, whose parity is even,
yields the coincidence
\begin{align}
C^{(1\,3\,5\,7)^{\top}}_{\bm{v}}
=\frac1{2^{7}}\left(1+\left|r_\sigma^{(4)}\right|\cos\psi_\sigma^{(4)}\right)
\end{align}
with~$\sigma=(1,2,4,3)$.
However, for the odd-parity output configurations in~$\xi^{(4)}$, such as~$\bm{\eta}=(1\,3\,5\,8)^{\top}$,
the coincidence is
\begin{align}
    C^{(1\,3\,5\,8)^{\top}}_{\bm{v}}=\frac1{2^{7}}\left(1-\left|r_\sigma^{(4)}\right|\cos\psi_\sigma^{(4)}\right).
\end{align}
If we continue to calculate the coincidence rates for all the rest of the output configurations in~$\xi^{(4)}$,
we find that output configurations
of the same parity give us the same coincidence rate.

To increase the counting rate, we sum over the coincidence rates of all the output configurations in~$\xi^{(4)}$ with even and odd parity,
respectively.
Then their differences are proportional to the cosine of the four-photon collective phase,
\begin{align}
    \sum_{\bm{\eta}\in\xi^{(4)}}C^{\bm{\eta}}_{\bm{v}}-\sum_{\bm{\eta}'\in\xi^{(4)}}C^{\bm{\eta}'}_{\bm{v}}=\frac1{2^{3}}\left|r_\sigma^{(4)}\right|\cos\psi_\sigma^{(4)}\label{eq:sum_coincidence_xi_four}
\end{align}
with~$\operatorname{par}(\bm{\eta})=-\operatorname{par}(\bm{\eta}')=1$.
With the pairwise overlaps~\eqref{eq:r4}
we get in Eq.~\eqref{eq:rrhoirhoi+1},
\begin{align}
    \left|r_\sigma^{(4)}\right|=\sqrt{r_{(1,2)}^{(4)}r_{(2,4)}^{(4)}r_{(4,3)}^{(4)}r_{(3,1)}^{(4)}}.
\end{align}
Then the four-photon collective phase can be calculated from Eq.~\eqref{eq:sum_coincidence_xi_four} to yield
\begin{equation}
\label{eq:psisigma4}
   \psi_{\sigma}^{(4)}=\arccos\left(\frac{2^{3}\left(\sum_{\bm{\eta}\in\xi^{(4)}}C^{\bm{\eta}}_{\bm{v}}-\sum_{\bm{\eta}'\in\xi^{(4)}}C^{\bm{\eta}'}_{\bm{v}}\right)}{\sqrt{r_{(1,2)}^{(4)}r_{(2,4)}^{(4)}r_{(4,3)}^{(4)}r_{(3,1)}^{(4)}}} \right).
\end{equation}

To recap, experimentally,
we need to collect four-photon coincidence data from the output configurations in Eq.~\eqref{eq:chi4} such that pairwise overlaps~\eqref{eq:r4} can be calculated from Eq.~\eqref{eq:rrhoirhoi+1}, 
and the four-photon coincidence data of the output configurations in $\xi^{(4)}$
to give the four-photon collective phase $\psi_{\sigma}^{(4)}$ from Eq.~\eqref{eq:psisigma4}.

Furthermore, for~$\bm{\eta}\in \xi^{(4)}$,
when we trace out one of the photons,
for example,
$\bm{\eta}=(1\,3\,5\,\bullet)^{\top}$,
we obtain
\begin{equation}
C^{(1\,3\,5\,\bullet)^{\top}}_{\bm{v}}=\sum_{i=1}^{8}C^{(1\,3\,5\,i)^{\top}},
\end{equation}
which includes two terms, $C^{(1\,3\,5\,7)^{\top}}_{\bm{v}}$ and~$C^{(1\,3\,5\,8)^{\top}}_{\bm{v}}$,
that are related to the four-photon collective phase.
The sum of these two terms is~$\nicefrac{1}{2^{6}}$ and is invariant with respect to the four-photon collective phase.
Thus, when~$\bm{\eta}\in\xi^{(4)}$,
we obtain genuine four-photon interference.

\subsection{\texorpdfstring{$n$ photons and~$2n$-mode interferometer}{n photons and 2n-mode interferometer}}
\label{subsec:multiphoton}
\begin{figure*}
\centering
    (a)~\includegraphics[width=0.36\columnwidth]{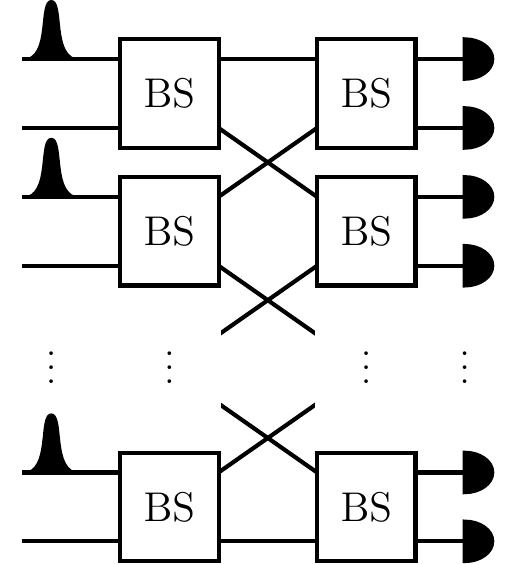}
    (b)~\includegraphics[width=0.3\columnwidth]{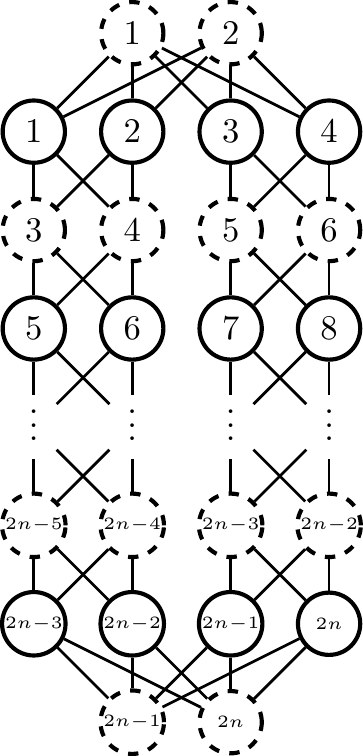}
    (c)~\includegraphics[width=0.3\columnwidth]{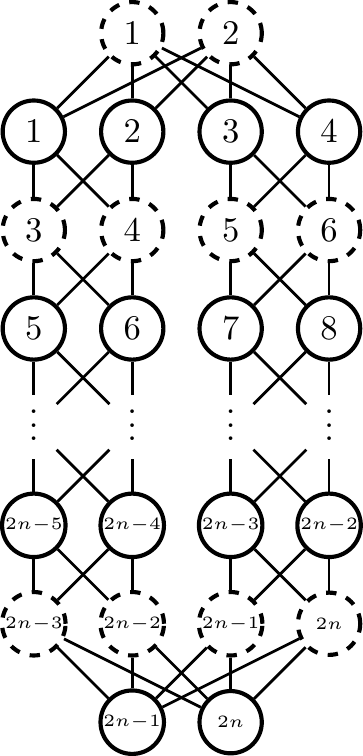}
    (d)~\includegraphics[width=0.15\columnwidth]{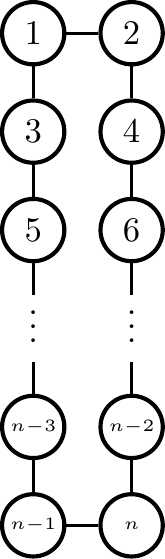}
    (e)~\includegraphics[width=0.15\columnwidth]{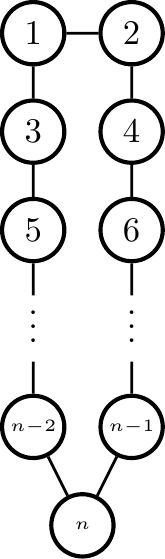}
    \caption{(a)~Generalized sparse interferometers for genuine~$n$-photon interference. 
    In the interferometer, there are only two layers of 50:50 BSs, with one photon injected into one input port of each BS in the first layer such that $\bm{v}=(1\,3\,\cdots\,2n-1)^{\top}$. For each second-layer BS,
    a single photon detector is placed at each output port.
    (b) and~(d) are the corresponding connectivity graph~$\mathsf{G}_{\text{c}}$ and enhanced-distinguishability graph~$\mathsf{G}_{\text{e}}$, respectively, for even~$n$.
    For odd~$n$,
    (c) and~(e) are the corresponding $\mathsf{G}_{\text{c}}$ and $\mathsf{G}_{\text{e}}$, respectively.}
    \label{fig:interferometern}
\end{figure*}

The sparse interferometer for the genuine four-photon interference in Fig.~\ref{fig:interferometerfour}(a) can be generalized to genuine $n$-photon interference with $2n$ input (or output) ports as shown in Fig.~\ref{fig:interferometern}(a).
Similarly to the four-photon case,
there are also two layers of BSs, with each layer comprising~$n$ BSs.
The $2n$ input ports of the interferometer are labeled by $1,2,\dots,2n$ from top to bottom.
The~$n$ photons are injected into the odd-number input port and a detector is placed at each output port of the interferometer.

We follow a similar procedure to that in Sec.~\ref{subsec:four-photon} to measure the $n$-photon collective phase.
In this case, $n$ photons are injected one into each odd-number input port of the interferometer.
We count coincidence events for all output configurations with $2n$ single-photon detectors.
Then the pairwise overlaps and the $n$-photon collective phase can be extracted from the coincidence rates of the output configurations in~$\chi^{(n)}_{\rho(i),\rho(i+1)}$ with $i\in[n]$ and in $\xi^{(n)}$ , respectively.
Here, $\rho$ is different 
for~$n$ being an odd number or an even number.
We provide a detailed explanation in the following.

For the generalized interferometer in Fig.~\ref{fig:interferometern}(a),
the corresponding scattering matrix can be written as
\begin{equation}
\label{eq:u2n}
   U=\frac12\begin{pmatrix}
        K&L&0&\cdots&\cdots&\cdots&0\\
        J&0&L&\ddots&\ddots&\ddots&\vdots\\
        0&\ddots&\ddots&\ddots&\ddots&\ddots&\vdots\\
        \vdots&\ddots&\ddots&\ddots&\ddots&\ddots&\vdots\\
        \vdots&\ddots&\ddots&\ddots&\ddots&\ddots&\vdots\\
        \vdots&\ddots&\ddots&\ddots&J&0&L\\
        0&\cdots&\cdots&\cdots&0&J&-K^{\top}
    \end{pmatrix}\in\text{U}(2n)
\end{equation}
and
\begin{align}
    J=\begin{pmatrix}
        1&1\\
        1&1
    \end{pmatrix},\,
    K=\begin{pmatrix}
        1&-1\\
        1&-1
    \end{pmatrix},\,
    L=\begin{pmatrix}
        -1&1\\
        1&-1
    \end{pmatrix}.
\end{align}
With the scattering matrix~\eqref{eq:u2n},
we can draw the corresponding connectivity graph $\mathsf{G}_\text{c}$ of the interferometer following the steps in Sec.~\ref{subsec:graphs}.

The connectivity graph $\mathsf{G}_{\text{c}}$ is slightly different for even~$n$ and odd $n$.
We give $\mathsf{G}_{\text{c}}$ for even~$n$ in Fig.~\ref{fig:interferometern}(b) and for odd~$n$ in Fig.~\ref{fig:interferometern}(c).
When the~$n$ photons are injected as the input-configuration vector
\begin{equation}
    \bm{v}=(1\,3\,5\,\cdots\,2n-1)^{\top},
\end{equation}
the enhanced-distinguishability graph $\mathsf{G}_{\text{e}}$ is shown in~Figs.~\ref{fig:interferometern}(d) and~\ref{fig:interferometern}(e) for even and odd~$n$, respectively.
Then
\begin{equation}
\label{eq:sigman}
    \sigma=\begin{cases}
        (1,2,4,6,\dots,n-2,n,n-1,\dots,5,3) & n \text{ is even}\\
        (1,2,4,6,\dots,n-1,n,n-2,\dots,5,3) & n \text{ is odd}
    \end{cases}
\end{equation}
is the $n$-length cycle in the enhanced-distinguishability graph.

Then we give the coincidence rates for the output configurations $\bm{\eta}\in\chi^{(n)}_{\rho(i),\rho(i+1)}$ with $\rho(i)$ given by~\eqref{eq:ncyclerho} and~\eqref{eq:sigman} for $i\in[n]$.
Note that each output configuration in
$\chi_{\rho(i),\rho(i+1)}^{(n)}$~\eqref{eq:chinrhoirhoi+1}
gives us the same coincidence rate,
namely,
\begin{align}
    C^{\bm{\eta}\in \chi_{\rho(i),\rho(i+1)}^{(n)}}_{\bm{v}}=\frac1{2^{2n-1}}\left(1-r_{(\rho(i),\rho(i+1))}^{(n)}\right)\label{eq:coincidence_chi_n}
\end{align}
and
\begin{equation}
\label{eq:numberoutputconfigschinrho}
\left|\chi_{\rho(i),\rho(i+1)}^{(n)}\right|
=\binom{n-1}{n-2}\cdot 2^{n-2}.
\end{equation}
Thus, when we sum over all the output configurations in~$\chi_{\rho(i),\rho(i+1)}^{(n)}$,
\begin{align}
    \sum_{\bm{\eta}\in \chi_{\rho(i),\rho(i+1)}^{(n)}}C^{\bm{\eta}}_{\bm{v}}=\frac{n-1}{2^{n+1}}\left(1-r_{(\rho(i),\rho(i+1))}^{(n)}\right),\label{eq:sum_coincidence_chi_n}
\end{align}
with~$i\in[n]$. 
From the summation of the coincidence rates of the output configurations in~$\chi^{(n)}_{\rho(i),\rho(i+1)}$,
we can extract the pairwise overlaps
\begin{equation}
\label{eq:rrhoirhoi+1n}
    r_{(\rho(i),\rho(i+1))}^{(n)}=1-\frac{2^{n+1}}{n-1}\sum_{\bm{\eta}\in\chi_{\rho(i),\rho(i+1)}^{(n)}}C_{\bm{v}}^{\bm{\eta}}\,
    \forall i\in[n]
\end{equation}
with
\begin{equation}
r_{(\rho(n),\rho(n+1))}^{(n)}\equiv r_{(\rho(n),\rho(1))}^{(n)}.
\end{equation}

For~$\bm{\eta}\in\xi^{(n)}$,
the coincidence is
\begin{align}
C^{\bm{\eta}}_{\bm{v}}=\frac1{2^{2n-1}}\left(1+(-1)^{\operatorname{par}(\bm{\eta})+n}|r_\sigma^{(n)}|\cos\psi_\sigma^{(n)}\right),\label{eq:coincidence_zeta_n}
\end{align}
with $\sigma$ in Eq.~\eqref{eq:sigman}.
Then the difference between coincidence rates of all the even- and odd-parity output configurations in~$\xi^{(n)}$ is
\begin{align}
\label{eq:sum_coincidence_zeta_n}
    \sum_{\bm{\eta}\in\xi^{(n)}}C^{\bm{\eta}}_{\bm{v}}-\sum_{\bm{\eta}'\in\xi^{(n)}}C^{\bm{\eta}'}_{\bm{v}}=\frac{(-1)^{n}}{2^{n-1}}|r_\sigma^{(n)}|\cos\psi_\sigma^{(n)}
\end{align}
with~$\operatorname{par}(\bm{\eta})=-\operatorname{par}(\bm{\eta}')=1$.
Notice that this is an exponentially small value,
which is an unavoidable feature of multiphoton interference.
Even for~$n$ photons interfering in an $n\times n$ Fourier interferometer,
which is conjectured to be optimal for the MCP in an $n\times n$ interferometer~(see the Appendix),
the weight for the $n$-photon collective-phase term is
\begin{equation}
\nicefrac{n!}{n^{n}}
    \approx \nicefrac{\sqrt{2\pi n}}{\mathrm{e}^{n}},
    \label{eq:nnn}
\end{equation}
which is also an exponentially small value.

With pairwise overlaps~\eqref{eq:pairwiseoverlap}
we obtain,
from Eq.~\eqref{eq:rrhoirhoi+1n},
the amplitude of the $n$-photon overlap in Eq.~\eqref{eq:sum_coincidence_zeta_n} being
\begin{align}
\label{eq:rsigman}
    |r_\sigma^{(n)}|=\sqrt{\prod_{i=1}^{n}r_{\rho(i),\rho(i+1)}^{(n)}}.
\end{align}
We substitute Eq.~\eqref{eq:rsigman} back into Eq.~\eqref{eq:sum_coincidence_zeta_n}
to get the $n$-photon collective phase
\begin{widetext}
\begin{equation}
\label{eq:psisigman}
    \psi_{\sigma}^{(n)}
    =\arccos\left(\frac{(-1)^{n}2^{n-1}\left(\sum_{\bm{\eta}\in\xi^{(n)}}C^{\bm{\eta}}_{\bm{v}}-\sum_{\bm{\eta}'\in\xi^{(n)}}C^{\bm{\eta}'}_{\bm{v}}\right)}{\sqrt{\prod_{i=1}^{n}r_{\rho(i),\rho(i+1)}^{(n)}}}\right)
\end{equation}
\end{widetext}
for $\sigma$ in Eq.~\eqref{eq:sigman}.
Here the collective phase cannot be uniquely determined
and there is an ambiguity in its sign because the coincidence rate is always an even function of the $n$-photon collective phase.

In experiments,
we need to collect coincidence data for output configurations in $\chi_{\rho(i),\rho(i+1)}^{(n)}$ to calculate the pairwise overlaps $r_{\rho(i),\rho(i+1)}^{(n)}$~\eqref{eq:rrhoirhoi+1n} for all $i\in[n]$.
Then together with the coincidence data of the output configurations in $\xi^{(n)}$,
we can estimate the $n$-photon collective phase~\eqref{eq:psisigman}.
Although from Eqs.~\eqref{eq:sum_coincidence_zeta_n} and~\eqref{eq:nnn} we see that the contribution of the highest-order collective phase is exponentially small, 
the total contributions of the $k$-photon collective phases 
with $k$ from 3 to $n$, 
which cannot be measured solely from two-photon HOM detection, 
can be pronounced in general multiphoton interference. 
Due to the scalability of our setup, 
it is feasible in our setup to measure the collective phases up to any order we want
and thus fully characterize multiphoton-interference behavior.

Moreover,
as we can see from the enhanced-distinguishability graphs in Figs.~\ref{fig:interferometern}(d) and~\ref{fig:interferometern}(e),
no lower-order MCP appears in our setup in principle. When
we trace out an arbitrary positive number of photons,
the $n$-photon collective phase is canceled out because there are an equal number of output configurations with odd and even parity, which gives us a coincidence rate with no dependence on the $n$-photon collective phase~\cite{Shchesnovich2018Phys.Rev.A} or any lower-order collective phase.
Thus, for~$\bm{\eta}\in\xi^{(n)}$, we indeed get genuine $n$-photon interference.

\section{Summary}
\label{sec:summary}
We have introduced a connectivity graph and an enhanced-distinguishability graph to help us design an interferometer with sparse structure suitable to generate genuine $n$-photon interference.
Our scheme involves doubling the number of interferometer modes,
but the sparsity of our interferometer trades off the expense of more modes against greatly reducing the optical depth.
Following the introduction of this interferometer design,
we further describe how to measure the $n$-photon collective phase via appropriate multiphoton coincidences
and explain how genuine $n$-photon interference is manifested in the multiphoton coincidence data for output configurations in $\xi^{(n)}$~\eqref{eq:chin}.

An explicit experimental setup is shown in Fig.~\ref{fig:interferometerfour}(a) for the four-photon case and in Fig.~\ref{fig:interferometern}(a) 
for the $n$-photon cases,
along with the details of the procedure for measuring four-photon and $n$-photon collective phases in Sec.~\ref{subsec:four-photon} and Sec.~\ref{subsec:multiphoton}, respectively.
In brief,
experimentally,
we record multiphoton coincidence events associated with the output configurations in
$\chi_{\rho(i),\rho(i+1)}^{(n)}$~\eqref{eq:chinrhoirhoi+1}
to give the pairwise overlaps of photons from Eq.~\eqref{eq:rrhoirhoi+1n},
which can be used to extract the $n$-photon collective phase from the recorded coincidence events of output configurations in~$\xi^{(n)}$ using Eq.~\eqref{eq:psisigman}.

In our proposal,
lower-order MCPs are eliminated by
setting scattering matrix elements to zero instead of previous techniques of manipulating photons' internal DoFs.
The advantage of our approach over the state of the art is that,
if we increase the photon number, we do not need to manipulate more photonic DoFs.
Furthermore,
the~$O(1)$ optical depth of our interferometer keeps the photon loss rate constant as the photon number increases.
Both of these advantages portend feasibility
for scaling up MCP to many photons.

\acknowledgments{
B.C.S.\ and J.W.\ are supported by the National Natural Science Foundation of China~(NSFC) with Grant No.~11675164
and Anhui Initiative in Quantum Information Technologies. 
J.W. would like to thank Rui Zhang and Sheng-Jun Yang for valuable discussions.
}

\appendix

\section{\texorpdfstring{Conjecture of optimal $n\times n$ interferometer for $n$-photon MCP}{Conjecture of optimal nxn interferometer for n-photon MCP}}
\label{app:conjecture}
Our conjecture is stated as follows.
Given~$n$ photons injected into an $n\times n$ interferometer with input and output configurations $\bm{v}=\bm{\eta}=(1\,2\,\cdots\,n)^\top$,
\begin{equation}
    F^{(n)}\in\operatorname*{arg\,max}_{U\in \mathrm{U}(n)} \left|\left(\bm{u}_{\bm{v}}^{\bm{\eta}}\right)^{\dagger}\cdot \Pi_\sigma\cdot \bm{u}_{\bm{v}}^{\bm{\eta}}\right|,\sigma=(1,2,\dots,n),
\end{equation}
where $F^{(n)}$ is an $n\times n$ Fourier matrix defined by
\begin{equation}
    \left(F^{(n)}\right)_{i,j}:=\frac{1}{\sqrt{n}}\mathrm{e}^{-\frac{2\pi\mathrm{i}}{n}(i-1)(j-1)},\forall i,j\in[n].
\end{equation}
To prove this conjecture,
we equivalently show
\begin{equation}
    \left|\left(\bm{u}_{\bm{v}}^{\bm{\eta}}\right)^{\dagger}\cdot \Pi_\sigma\cdot \bm{u}_{\bm{v}}^{\bm{\eta}}\right|\leq \nicefrac{n!}{n^n},\,\forall U\in\mathrm{U}(n),\sigma=(1,2,\dots,n)
\end{equation}
or,
more compactly,
\begin{equation}
    \left|\operatorname{perm}\left(U^{*}\circ U_{\sigma}\right)\right|\leq \nicefrac{n!}{n^n},\,\forall U\in\mathrm{U}(n),\sigma=(1,2,\dots,n),
\end{equation}
where $\circ$ is the elementwise product of matrices
and $U_\sigma$ is the matrix from $U$ by permuting its column by $\sigma$,
i.e., $\left(U_{\sigma}\right)_{i,j}:=U_{i,\sigma(j)}$.

Here, our proof is for $n=3$.
We need to use the result
that,
for an arbitrary $n\times n$ complex matrix $V$~\cite{Carlen2006MethodsAppl.Anal.},
\begin{equation}
    \left|\operatorname{perm}V\right|\leq n!\left(\prod_{i=1}^{n}\sum_{j=1}^{n}\frac{\left|V_{i,j}\right|^{2}}{n}\right)^{\frac{1}{2}}.
\end{equation}
Substituting
$V=U^{*}\circ U_{\sigma}$, we have
\begin{equation}
    \left|\operatorname{perm}(U^{*}\circ U_{\pi})\right|\leq 3!\left(\prod_{i=1}^{3}\sum_{j=1}^{3}\frac{1}{3}\left|U_{i,j}U_{i,\sigma(j)}^{*}\right|^{2}\right)^{\nicefrac{1}{2}}.\label{eq:uupi}
\end{equation}
Notice that
\begin{align}
    &|U_{i,1}|^2|U_{i,2}^{*}|^2+|U_{i,2}|^2|U_{i,3}^{*}|^2+|U_{i,3}U_{i,1}^{*}|^2\\
    &\leq\frac{1}{3}\left(|U_{i,1}|^2+|U_{i,2}|^2+|U_{i,3}|^2\right)^2=\nicefrac{1}{3}\label{eq:ui1plus}
\end{align}
where we have used the inequality
\begin{equation}
    ab+bc+cd\leq \frac{1}{3}(a+b+c)^2,\forall a,b,c\in \mathbb{R}
\end{equation}
and the unitarity of $U$.
Taking Eq.~\eqref{eq:ui1plus} back to Eq.~\eqref{eq:uupi},
we obtain
\begin{equation}
    \left|\operatorname{perm}(U\circ U_{\sigma})\right|\leq 3!\left(\prod_{i=1}^{3}\nicefrac{1}{3^2}\right)^{\nicefrac{1}{2}}=\nicefrac{3!}{3^3},
\end{equation}
which finishes the proof for $n=3$.

\bibliography{collective_phase}
\end{document}